\documentclass{article}
\usepackage{icrctc07}

\def \yr {\mathrm{\ year}}

\newcommand{\degree}{$^\circ$}
\newcommand{\degrees}{\degree}

\title{Measurement of the UHECR
  spectrum above \protect$10^{19}$~eV at the Pierre Auger Observatory using
  showers with zenith angles greater than 60\protect\degree}
\shorttitle{UHECR spectrum at the Auger Observatory using showers
  with zenith  greater than 60\protect\degree}
\authors{P. Facal San Luis$^{1}$, for the Pierre Auger Collaboration$^{2}$}
\shortauthors{P. Facal San Luis, for the Pierre Auger Collaboration}
\afiliations{$^{1}$Universidade de Santiago de Compstela and IGFAE, Campus Sur, 15782, Santiago, Spain\\
$^{2}$Observatorio Pierre Auger, Av. San Mart´n Norte 304, (5613) Malarg\"{u}e, Mendoza, Argentina}

\email{facal@fpaxp1.usc.es}

\abstract{ We report a measurement of the cosmic ray energy spectrum obtained
  using the inclined events detected with the Pierre Auger
  Observatory. Showers with zenith angles between 60$^\circ$ and
  80$^\circ$ recorded in the period between 1 January 2004 and 28
   February 2007 are analysed. Showers are first reconstructed in
  arrival direction and then fitted to density maps of the muon
  numbers obtained from $10^{19}$eV simulated proton showers for
  different arrival directions, in order to obtain the core position and an
  overall normalisation factor $N_{19}$ which is used as an energy
  estimator. The parameter $N_{19}$ is shown to be correlated with the
  shower energy measured with the fluorescence technique for a
  sub-sample of good quality hybrid showers.
 This correlation, measured with hybrid events, is then used to
  determine the energy of all the showers.  }

\begin{document}
\maketitle
\section{Introduction}
Inclined showers are detected regularly  with the Pierre Auger Observatory. 
They are of interest because they enhance both the exposure of 
the detector and its sky coverage. Showers induced by hadronic nuclei 
with zenith angles greater than 60$^\circ$ are mainly composed of muons at 
ground level and their detection provides 
complementary information, relevant for composition and hadronic 
model studies. In addition inclined events constitute a background for the 
detection of neutrino-induced showers.

The Pierre Auger Observatory combines the surface and fluorescence
techniques to study high-energy cosmic ray showers. The surface
detector (SD), described in~\cite{Abraham:2004dt}, uses 1.2~m deep
water-Cherenkov tanks that provide enhanced sensitivity to muons and
make the Auger Observatory suitable for studying inclined showers.
Inclined events are reconstructed using a special analysis procedure
to account for the muons deviating in the geomagnetic
field~\cite{poster}.  The energy assignment is performed using an
estimator that is calibrated with a subset of events (hybrid)
which are also detected with the fluorescence detector (FD), 
in a manner similar to what is done for showers below 60\degree.
The energy spectrum of cosmic ray above 6.3~EeV and with zenith angles
between 60\degree~and 80\degree~as measured with the Pierre Auger
Observatory is presented for the first time and shown to be consistent
with that measured for events below 60\degree.

\section{Analysis and results}
The event reconstruction is essentially a two-fold process. First the
arrival direction of the shower is reconstructed using the measured
start time of the signals in the tanks. Then the core and the size of
the shower are determined using the relative distributions of the muon
number densities at ground level, ``muon maps'', which are obtained
from simulation ~\cite{poster}. 
The muons entering the tank are converted to signal by convoluting with the 
tank response which has been simulated with the GEANT4 package, 
accounting for the different relevant processes. 
Signal probability distributions are evaluated and the best core position
and $N_{19}$, the normalization factor of the muon map,   are obtained using a
maximum likelihood method.  $N_{19}$ can be used as an energy
estimator and its relation to shower energy is determined  experimentally
using inclined hybrid events.

Before comparing the measured signal to the muon maps, the
electromagnetic component of the signal is subtracted. 
Close to $60^\circ$ an electromagnetic contribution from the main
showering process, arising from neutral pions, 
can be expected, particularly close to the shower
axis. At very high zenith angles the only electromagnetic contribution
arises from the muons themselves, mainly through muon decay in flight,
and is of order 15$\%$.
These contributions have been calculated using simulations of protons
with AIRES at different energies and zenith angles (proton primaries,
being more penetrating, have the largest electromagnetic
contribution).
The fraction of electromagnetic signal
to the total has been parametrised as a function of zenith angle and
distance to the core.

\begin{figure}[t]
\centering
\includegraphics*[width=0.45\textwidth,angle=0,clip]{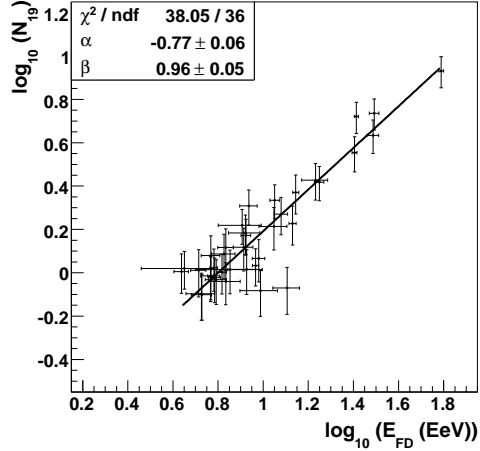}
\caption{\label{figcalfd} Correlation between FD energies and
  \protect$N_{19}$ in double logarithmic scale. The calibration curve,
  obtained by a linear fit to the data points,  is shown superimposed.}
\end{figure}

The reconstructed $N_{19}$ values are calibrated using the sub-sample
of events which are  recorded simultaneously by the FD.  
For these hybrid events a direct measurement of the energy
released in the atmosphere by the electromagnetic component of the
shower is available. The yield used to estimate the FD energy is taken from~\cite{yield}.
 The events used are selected according to a set
of standard quality cuts in the FD
reconstruction~\cite{HybridQC}, with minor adjustments optimised for
this analysis.  The requirement that the shower
maximum is well contained in the field of view of the FD strongly constrains the
geometry of inclined events and there are no events above 75\degrees.
The correlation between FD energies and
$N_{19}$ values is shown in figure~\ref{figcalfd}.  The calibration curve
is obtained by a linear fit to the data points in this logarithmic
plot, in the form $N_{19} = 10^{\alpha} E_{FD}^{\beta}$,
that yields best fit values of $\alpha=-0.77 \pm 0.06$ and $\beta= 0.96\pm 0.05$.

\begin{figure}[t]
\centering
\includegraphics*[width=0.4\textwidth,angle=0,clip]{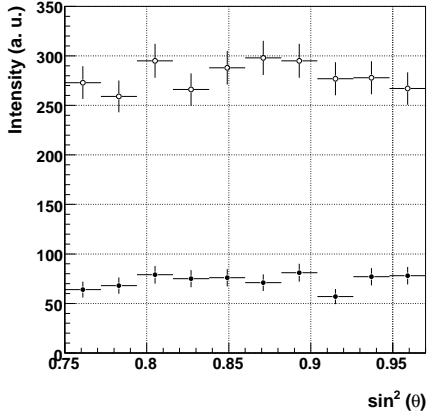}
\caption{\label{figsin2} Distribution of \protect$\sin^{2}\theta$
  for T5 events and two different values of \protect$N_{19}$
  (\protect$1.0> N_{19}> 0.4$, open circles; \protect$N_{19}>1.0$, full
  circles). The distribution flattens for higher $N_{19}$ as the
  detector reaches full efficiency.}
\end{figure}

A high-level trigger (T5) is defined for the SD events;
it has a two-fold purpose, to assure the quality of the reconstruction
avoiding events falling close to the border of the array and to allow
a simple geometrical calculation of the exposure. 
 The T5 definition requires that the tank closest to the reconstructed core
is surrounded by an hexagonal ring of working stations.
 With this definition, the aperture is calculated, for a
given array configuration, by counting the number of T5 hexagons and
integrating in solid angle.  The aperture for events with zenith
angles exceeding $80^\circ$ only represents  about 12\% of that above
$60^\circ$ and these events are discarded because as the zenith angle
increases the uncertainty associated with the angular reconstruction
rises.
In addition, for zenith angles above $80^\circ$  the triggering
efficiency decreases rapidly.  The total exposure is determined by
integrating the instantaneous aperture weighted by  the detection efficiency
over the different array configurations during the period of time. The
detection efficiency has been calculated using the muon maps.  For
values of $N_{19}>1$ the efficiency integrated over the solid angle
range exceeds  98\%.  Only events with $N_{19}>1$ ($E \sim
6.3$~EeV) are considered for the present analysis.
In figure~\ref{figsin2} the $\sin^2 \theta$-distribution is shown to be
flat for $N_{19}>1$ consistent with the result deduced from
simulation.

\begin{figure}[t]
\centering
\includegraphics*[width=0.48\textwidth,angle=0,clip]{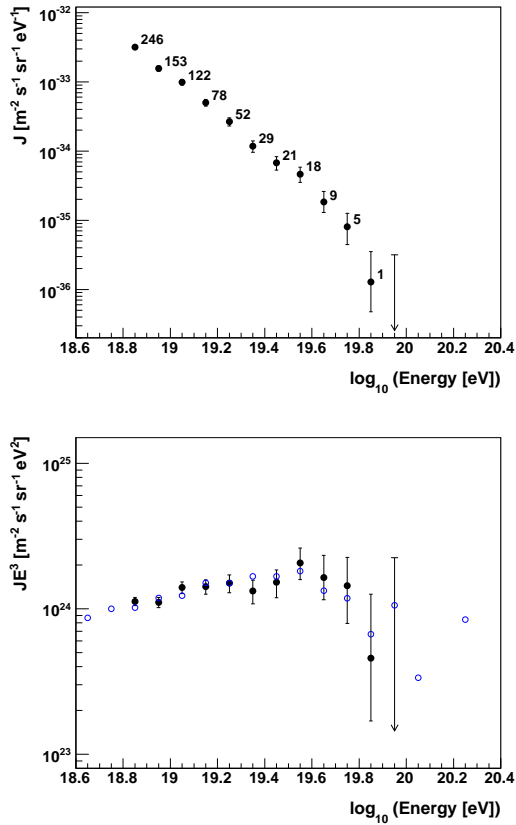}
\caption{\label{figspec} 
  Upper panel, inclined event energy spectrum (statistical
  errors or 95\% CL, number of events in each bin indicated) . 
  Lower panel, spectrum multplied by $E^{3}$. 
  The spectrum obtained for events below 60\protect\degree~\protect\cite{VS} is superimposed 
  (blue open circles).}
\end{figure}

\begin{figure}[t]
\centering
\includegraphics*[width=0.42\textwidth,angle=0,clip]{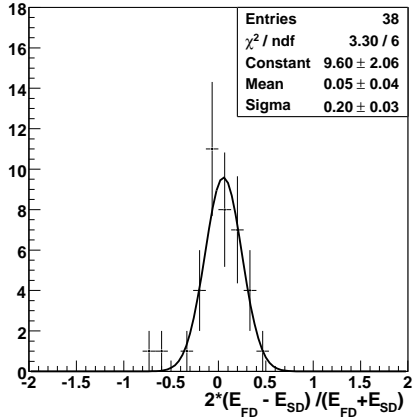}
\caption{\label{fighistofd} Relative difference between FD energies
  and calibrated energies for the events in the calibration plot. }
\end{figure}

The cosmic ray energy spectrum in the angular range between $60^\circ$
and $80^\circ$ as measured by the Pierre Auger observatory between
1 January 2004 and 28 February 2007 is shown in
figure~\ref{figspec}. 
A total of 734  events are used to build the spectrum and the
integrated exposure in the period amounts to
1510~${\mathrm {\ km^2 sr \yr}}$, i.e. $29\%$ of the
exposure for events below 60\degree~\cite{VS}.

\section{Discussion}
The reconstruction of inclined showers is a relatively new challenge.
Some of the uncertainties in the reconstruction are avoided by using
the FD energy calibration as for the analysis for showers below 60\degree.
There are still a number of systematic uncertainties which need to be
discussed in some detail.  Several test and cross-checks have been
performed to ensure the validity of the results, as discussed below. 

The inclined shower reconstruction uses simulated muon maps. For a
given arrival direction the shape of the muon distributions is quite
insensitive to the energy, to the composition and to the hadronic
model used in the simulation~\cite{ave}. Differences can be quantified
by an overall normalization.  The procedure to obtain the energy by
correlating $N_{19}$ with the FD energy takes care of a great part of
the systematic normalization changes. The maps implicitly account for the
attenuation of the muon content due to the different amounts of matter
traversed for the different zenith angles.  The study of the $\sin^2
\theta$ distribution (figure~\ref{figsin2}) suggests that this effect
is below the current level of statistical uncertainties.  There is a
systematic uncertainty which stems from the angular uncertainty, of
order 1\degree, in the reconstruction which translates directly to a
change of $N_{19}$. The corresponding uncertainty in $N_{19}$ increases 
as the zenith angle rises and has a maximum value of $12\%$ at $80^\circ$.

The reconstruction process depends on the models and on composition,
mainly through the electromagnetic component, introducing possibly the
largest systematic uncertainty. The effect has been explored by
changing the fraction of electromagnetic correction applied to the
data by an overall normalization factor that ranges between 1.5 and 0.5.  The net
effect on average is an overall change in $N_{19}$ for showers of
zenith angle above $65^\circ$ that is independent of zenith angle and
energy. Such systematic change would be on average reabsorbed in the
energy correlation plot. Below $65^\circ$ the average normalization
obtained in the reconstruction shifts by less than $7\%$.  We
tentatively assign this value to the systematic error associated to the
electromagnetic part.

The calibration curve is another possible source of systematic
uncertainty. Effects due to the cuts applied to assure the quality
of the reconstruction of the FD events have been
carefully evaluated and are at the level of 10\%, within the
statistical significance of the calibration curve. Also, currently only 38 events 
are available.

The uncertainty in the measurements of the aperture is $\sim 3\%$,
negligible in view of other uncertainties. The effect of the quality
trigger has been evaluated using well contained showers found in the
data and randomly repositioning them in the real array. The
distribution of the $N_{19}$ values obtained after the reconstruction
of the events has an RMS value of $7\%$. The fraction of events that
are misreconstructed to be outside the array is $4\%$.

At the moment the main source of systematic uncertainty in the
analysis comes directly form the uncertainty in the FD energy scale,
that is quoted at 22\% level, dominated by a $14~\%$ uncertainty in
the fluorescence yield measurement, $11\%$ in the detector
calibration and $10\%$ in the reconstruction method~\cite{FDsystem}.

The hybrid events can be used also to test the SD
reconstruction. The distribution of the difference between the SD and
the FD hybrid-reconstructed energies normalised to the FD energy is
shown in figure~\ref{fighistofd}.  The RMS fractional deviation is
$(20\pm3)\%$.  This is consistent with the combination of statistical
uncertainties, uncorrelated FD systematics uncertainties, shower to
shower fluctuations and the systematic uncertainties estimated 
for this analysis.

The spectrum observed is in good agreement with the SD 
spectrum for events below 60\degree. 
The comparison between the two spectra has
implications for composition and/or hadronic models. This is
presently under study.


\begin{thebibliography}{99}
\bibitem{Abraham:2004dt} J.\ Abraham {\it et al.}  [P. Auger
  Collaboration], Nucl.\ Inst.\ Meth.\  {\bf A523} (2004) 50.

\bibitem{poster} D.\ Newton [P. Auger Collaboration], these
  proceedings \#0308.

\bibitem{yield}M.\ Nagano et al.,   Astrop.\ Phys.\ {\bf 22} (2004) 235. 

\bibitem{HybridQC} L.\ Perrone [P. Auger Collaboration], these
  proceedings \#0316.

\bibitem{VS} M.\ Roth [P. Auger Collaboration], these proceedings \#0313.

\bibitem{ave} M.\ Ave et al.,
  Astrop.\ Phys.\ {\bf 14} (2000) 91. 

\bibitem{FDsystem} B. Dawson [P. Auger Collaboration], these proceedings \#0976.

\end{thebibliography}
\end{document}